\documentclass{emulateapj}

\usepackage{amssymb}

\def\ts     {\thinspace}
\def\kms    {\ts km\ts s$^{-1}$}
\def\etal   {{\rm et\ts al.}}

\def\aco    {{\rm CO}($J$=1$\to$0)}
\def\acoalt {{\rm CO} $J$=1$\to$0}

\def\sio    {{\rm SiO}($J$=12$\to$11)}
\def\ccn    {{\rm CN}($N$=3$\to$2)}
\def\dcn    {{\rm CN}($N$=4$\to$3)}

\def\ehcn    {{\rm HCN}($J$=5$\to$4)}
\def\fhcn    {{\rm HCN}($J$=6$\to$5)}
\def\ehnc    {{\rm HNC}($J$=5$\to$4)}
\def\fhnc    {{\rm HNC}($J$=6$\to$5)}

\def\ehco    {{\rm HCO$^+$}($J$=5$\to$4)}
\def\fhco    {{\rm HCO$^+$}($J$=6$\to$5)}
\def\altehcn    {{\rm HCN} $J$=5$\to$4}
\def\fhncalt    {{\rm HNC} $J$=6$\to$5}
\def\bhcccn    {{\rm HC$_3$N}($J$=49$\to$48)}
\def\chcccn    {{\rm HC$_3$N}($J$=57$\to$56)}
\def\hcccn    {{\rm HC$_3$N}($J$=59$\to$58)}
\def\dhcccn    {{\rm HC$_3$N}($J$=60$\to$59)}

\def\apm    {APM\,08279+5255}
\slugcomment{draft version \today, accepted for publication in the Astrophysical Journal}

\shorttitle{HCN, HNC, and \fhco\ in \apm\ ($z$=3.91)}
\shortauthors{Riechers et al.}

\begin{document}

\title{
Dense Molecular Gas Excitation in Nuclear Starbursts at High Redshift: \\
HCN, HNC, and \fhco\ Emission in the $z$=3.91 Quasar Host of APM\,08279+5255}

\author{Dominik A. Riechers\altaffilmark{1,6}, Axel Wei\ss\altaffilmark{2}, 
Fabian Walter\altaffilmark{3}, and Jeff Wagg\altaffilmark{4,5}}

\altaffiltext{1}{Astronomy Department, California Institute of
  Technology, MC 249-17, 1200 East California Boulevard, Pasadena, CA, USA
  91125; dr@caltech.edu}

\altaffiltext{2}{Max-Planck-Institut f\"ur Radioastronomie, Auf dem
  H\"ugel 69, Bonn, D-53121, Germany}

\altaffiltext{3}{Max-Planck-Institut f\"ur Astronomie, K\"onigstuhl 17, 
  Heidelberg, D-69117, Germany}

\altaffiltext{4}{National Radio Astronomy Observatory, PO Box O, Socorro, NM 87801,USA}

\altaffiltext{5}{European Southern Observatory, Alonso de C{\'o}rdova 3107, Vitacura, Casilla 19001, Santiago 19, Chile}

\altaffiltext{6}{Hubble Fellow}

%\email{dr@caltech.edu}

\begin{abstract}

We report the detection of surprisingly strong \fhcn, \fhnc, and
\fhco\ emission in the host galaxy of the $z$=3.91 quasar \apm\
through observations with the Combined Array for Research in
Millimeter-wave Astronomy (CARMA). HCN, HNC, and HCO$^+$ are typically
used as star formation indicators, tracing dense molecular hydrogen
gas [$n$(H$_2$) $> 10^5$\,cm$^{-3}$] within star-forming molecular
clouds. However, the strength of their respective line emission in the
$J$=6$\to$5 transitions in \apm\ is extremely high, suggesting that
they are excited by another mechanism besides collisions in the dense
molecular gas phase alone.  We derive $J$=6$\to$5 line luminosities of
$L'_{\rm HCN} = (4.9 \pm 0.6)$, $L'_{\rm HNC} = (2.4 \pm 0.7)$, and
$L'_{\rm HCO^+} = (3.0 \pm 0.6) \times 10^{10}$\,$\mu_{\rm
L}^{-1}$\,K\,\kms\,pc$^2$ (where $\mu_{\rm L}$ is the lensing
magnification factor), corresponding to $L'$ ratios of
$\sim$0.23--0.46 relative to \aco.  Such high line ratios would be
unusual even in the respective ground-state ($J$=1$\to$0) transitions,
and indicate exceptional, collisionally and radiatively driven
excitation conditions in the dense, star-forming molecular gas in
\apm.  Through an expansion of our previous modeling of the HCN line
excitation in this source, we show that the high rotational line
fluxes are caused by substantial infrared pumping at moderate
opacities in a $\sim$220\,K warm gas and dust component. This implies
that standard $M_{\rm dense}$/$L'$ conversion factors would
substantially overpredict the dense molecular gas mass $M_{\rm
dense}$.  We also find a \fhcn/\ehcn\ $L'$ ratio greater than 1
(1.36$\pm$0.31) -- however, our models show that the excitation is
likely not `super-thermal', but that the high line ratio is due to a
rising optical depth between both transitions. These findings are
consistent with the picture that the bulk of the gas and dust in this
source is situated in a compact, nuclear starburst, where both the
highly active galactic nucleus and star formation contribute to the
heating.

\end{abstract}


\section{Introduction}

Over the past decade, great progress has been made in understanding
the conditions for star formation in gas-rich galaxies out to the
highest redshifts. Molecular gas, the prospective fuel for star
formation, is now detected in more than 70 galaxies at $z>1$, allowing
to compare the properties of star-forming environments among different
galaxy populations in the early universe (see Solomon \& Vanden Bout
\citeyear{sv05} for a review).  These detections are almost
exclusively being obtained in rotational transitions of CO, which (due
to its relatively low critical density of $n_{\rm crit}$(H$_2$)
$\simeq 300$\,cm$^{-3}$) is a good proxy for the total amount of
molecular gas in a galaxy.

More focused studies of the {\em dense} molecular gas found in the
star-forming cores of molecular clouds typically employ observations
of high dipole moment, high critical density ($n_{\rm crit}$(H$_2$) $>
10^4$\,cm$^{-3}$) molecules such as HCN, HCO$^+$, and HNC, both in
nearby galaxies and out to high $z$ (e.g., Gao \& Solomon
\citeyear{gs04}; Riechers et al.\ \citeyear{rie06a},
\citeyear{rie07a}; Gao et al.\ \citeyear{gao07}; 
Baan et al.\ \citeyear{baa08}; Gracia-Carpio et al.\
\citeyear{gra08}).

We here aim to study, for the first time, the dense, star-forming
molecular gas {\em excitation} in a high-$z$ galaxy. Constraints on
the excitation of gas at very high densities and the physical
mechanisms (collisions vs.\ other channels) responsible are crucial to
understand in more detail how excitation may influence scaling
relations between the dense gas content and star formation rate of
galaxies back to early cosmic times. To disentangle excitation effects
from other phenomena (such as the chemical composition of the gas), it
is crucial to target multiple diagnostic lines in a well-studied, key
system. The target of this study is the extremely luminous $z$=3.91
quasar \apm, which has been studied comprehensively in CO and other
diagnostics (see, e.g., Wei\ss\ et al.\ \citeyear{wei07}; Riechers et
al.\ \citeyear{rie09a} for details).  In particular, it is one out of
only two high-$z$ galaxies in which multiple dense molecular gas
tracers were detected to date (the other being the Cloverleaf quasar
at $z$=2.56; Barvainis et al.\ \citeyear{bar97}; Solomon et al.\
\citeyear{sol03}; Wagg et al.\ \citeyear{wag05}; Riechers et al.\
\citeyear{rie06a}, \citeyear{rie07}; Garcia-Burillo et al.\ 
\citeyear{gar06}; Guelin et al.\ \citeyear{gue07}).

\begin{figure*}
\epsscale{1.15}
\plotone{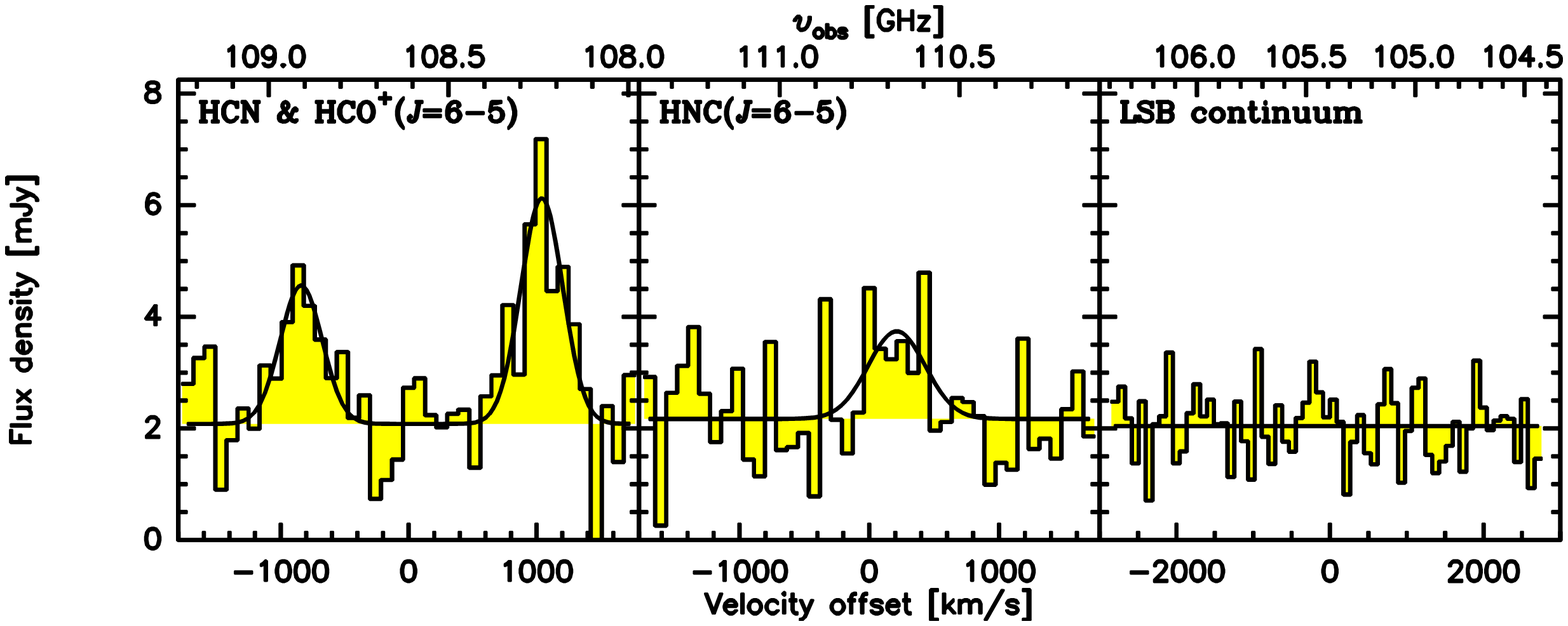}
%\vspace{-10mm}

\caption{CARMA spectra of the \fhcn, \fhco\ ({\em left}), \fhnc\ 
({\em middle}), and continuum emission at 2.76, 2.71, and 2.84\,mm
({\em left} to {\em right}) toward \apm\ at a resolution of
$\sim$85\,\kms\ (31.25\,MHz).  The velocity scales are relative to the
tuning frequencies of 108.611\,GHz ({\em left}) and 110.751\,GHz ({\em
middle}) and the central frequency of both LSBs at 105.391\,GHz ({\em
right}). The rms per velocity bin are 0.95, 0.88, and
$\sim$0.65--0.92\,mJy ({\em left} to {\em right}). The solid curves
show simultaneous Gaussian fits to the line and continuum emission
where applicable.
\label{f1}}
%\vspace{-5mm}
%
\end{figure*}

In this paper, we report the detection of \fhcn, \fhco, and \fhnc\
emission in the quasar host galaxy of \apm\ ($z$=3.91), using the
Combined Array for Research in Millimeter-wave Astronomy
(CARMA). These observations represent the first extragalactic
detections of such high-$J$ lines of the dense gas tracers HCN,
HCO$^+$, and HNC, and significantly constrain the physical properties
of the dense gas in the star-forming regions of this distant
galaxy. We use a concordance, flat $\Lambda$CDM cosmology throughout,
with $H_0$=71\,\kms\,Mpc$^{-1}$, $\Omega_{\rm M}$=0.27, and
$\Omega_{\Lambda}$=0.73 (Spergel \etal\ \citeyear{spe03},
\citeyear{spe07}).

\section{Observations}

We used CARMA to observe the \fhcn\ ($\nu_{\rm rest}$ =
531.7164\,GHz), \fhco\ (535.0618\,GHz), and \fhnc\ (543.8974\,GHz)
transition lines ($\sim$560\,$\mu$m) toward \apm.  At $z$=3.911, these
lines are redshifted to 108.270, 108.952, and 110.751\,GHz
($\sim$2.8\,mm).  The target was observed with 14 or 15 antennas
(corresponding to 91 or 105 baselines per antenna configuration) for
14\,tracks in March and June 2008 (setup 1; HCN/HCO$^+$, observed
simultaneously) and 23\,tracks between February and July 2009 (setup
2; HNC), amounting to a total observing time of 148\,hr. This results
in 37\,hr on source time for HCN/HCO$^+$ and 52\,hr for HNC. All
HCN/HCO$^+$ observations were carried out in D array (11--148\,m
baselines), while HNC observations were carried out in the C, D and E
arrays (6--185\,m baselines after flagging). Observations before June
2008 were carried out with the previous generation 3\,mm receivers,
and observations after March 2008 were carried out with the new
generation 3\,mm receivers (which offer improved noise temperatures,
tuning range, and stability).

Weather conditions scaled between acceptable and excellent for
observations at 3\,mm wavelengths. For the HCN/HCO$^+$ observations,
typical median phase rms values were 275\,$\mu$m (normalized to a
100\,m baseline, measured at 45$^\circ$ elevation), median optical
depths were $\tau_{\rm 230\,GHz}$=0.24, and median precipitable water
vapor columns were 3.9\,mm. For the HNC observations, typical median
phase rms values were 300 (C array)\footnote{Only one 3\,hr track was
observed in C array, and all baselines $>$185\,m were discarded due to
the relatively high phase noise.}, 235 (D array), and 395\,$\mu$m (E
array), median optical depths were $\tau_{\rm 230\,GHz}$=0.38, and
median precipitable water vapor columns were 5.5\,mm.  The nearby
source J0818+423 (distance to \apm: 10.6$^\circ$) was observed every
20 minutes for secondary amplitude and phase calibration. The strong
calibrator sources J0423-013, J0927+390, 3C\,111, 3C\,84, and 3C\,273
were observed at least once per track for bandpass and secondary flux
calibration. Absolute fluxes were bootstrapped relative to Mars,
Uranus, or 3C\,84 (when no planet was available).  Pointing was
performed at least every 2-4\,hr on nearby sources, using both radio
and optical modes.  The resulting total calibration is estimated to be
accurate within $\sim$15\% (see Appendix).

The 3\,mm receivers were tuned between the HCN and HCO$^+$ lines at
108.611\,GHz (setup 1) and at the redshifted HNC frequency of
110.751\,GHz (setup 2), both in the upper sideband (USB). The
intermediate frequencies were chosen to be 1.79\,GHz for the
HCN/HCO$^+$ observations (optimized for the previous generation
receivers) and 2.50\,GHz for the HNC observations (optimized for the
new generation receivers), centering the lower sidebands (LSBs) at
105.031 and 105.751\,GHz. Three bands with 15 channels of 31.25\,MHz
($\sim$85\,\kms ) width each were centered on the tuning
frequencies. The bands were overlapped by 2 channels to improve
calibration of the correlated dataset, leading to an effective
bandwidth of 1281.25\,MHz ($\sim$3500\,\kms ) per sideband. This
comfortably covers the HCN, HCO$^+$, and HNC emission lines as well as
the 2.8\,mm (rest-frame 560\,$\mu$m) continuum of the source. It also
provides coverage of the redshifted HC$_3$N $J$=57$\to$56, 59$\to$58,
and 60$\to$59 lines ($\nu_{\rm rest}$ = 518.1897, 536.3420, and
545.4170\,GHz; $\nu_{\rm obs}$ = 105.516, 109.212, and 111.060\,GHz),
as well as the \sio\ line ($\nu_{\rm rest}$ = 520.8782\,GHz; $\nu_{\rm
obs}$ = 106.064\,GHz). In particular, \chcccn\ lies in a spectral
range that is covered by both LSB frequency setups.

For data reduction and analysis, the MIRIAD package was used. The
final plots were created with the GILDAS package. All data were imaged
using `natural' weighting, resulting in synthesized clean beams of
4.5$''$$\times$3.6$''$ (HCN/HCO$^+$), and 5.3$''$$\times$4.6$''$ (HNC;
$\sim$29 and 35\,kpc at $z$=3.91). The CO size of the source is
$<$0.4$''$ (Riechers et al.\ \citeyear{rie09a}), i.e., it is
unresolved at the resolution of our observations. The final rms values
for the HCN/HCO$^+$ and HNC observations are 0.15 and
0.13\,mJy\,beam$^{-1}$ over the full USB bandpass (1281.25\,MHz), and
0.95 and 0.88\,mJy\,beam$^{-1}$ per 31.25\,MHz ($\sim$85\,\kms )
channel. The rms noise in the LSB is by 3\% (HCN/HCO$^+$) and 15\%
(HNC) better compared to the USB due to atmospheric transmission.

%%%%%%%%%%%%%%%%%%%%%%%%%%%%%%%%%%%%%%%%%%%%%%%%
%%%% Tab.1: Line Parameters
%%%%%%%%%%%%%%%%%%%%%%%%%%%%%%%%%%%%%%%%%%%%%%%%
\begin{deluxetable}{ l c c}
\tabletypesize{\scriptsize}
\tablecaption{Measured line fluxes and luminosities in \apm. \label{t1}}
\tablehead{
& $ I $ & $L'$ \\
& [Jy\,\kms\ ] & [10$^{10}$\,$\mu_{\rm L}^{-1}$\,K\,\kms\,pc$^2$] }
\startdata
\fhco\   & 1.01 $\pm$ 0.19 & 3.0 $\pm$ 0.6  \\
\fhcn\   & 1.65 $\pm$ 0.19 & 4.9 $\pm$ 0.6 \\
\fhnc\   & 0.86 $\pm$ 0.24 & 2.4 $\pm$ 0.7  \\
\chcccn\tablenotemark{a} & $<$0.34	   & $<$1.0	  \\
\hcccn\  & $<$0.52	   & $<$1.5	  \\
\dhcccn\ & $<$0.49	   & $<$1.4	  \\
\sio\    & $<$0.44	   & $<$1.3	  %\\
\vspace{-1mm}
\enddata 
\tablecomments{${}$
All limits are 3$\sigma$, extracted over a linewidth of 400\,\kms.
Luminosities are derived as described by Solomon \etal\
(\citeyear{sol92}): $L'[{\rm K
\ts km\ts s^{-1} pc^2}] = 3.25 \times 10^7 \times I \times \nu_{\rm
obs}^{-2} \times D_{\rm L}^2 \times (1+z)^{-3}$, where $I$ is the
velocity--integrated line flux in Jy
\kms, $D_{\rm L}$ is the luminosity distance in Mpc ($z=3.911$), and
$\nu_{\rm obs}$ is the observed frequency in GHz.  The given
luminosities are not corrected for the lensing magnification factor of
$\mu_{\rm L}$=4.2 (Riechers et al.\ \citeyear{rie09a}).
\tablenotetext{a}{Covered by both frequency settings.}
}
\end{deluxetable}

%%%%%%%%%%%%%%%%%%%%%%%%%%%%%%%%%%%%%%%%%%%%%%%%%%%%%%%%%%%

\section{Results}

\begin{figure*}
\epsscale{1.15}
\plotone{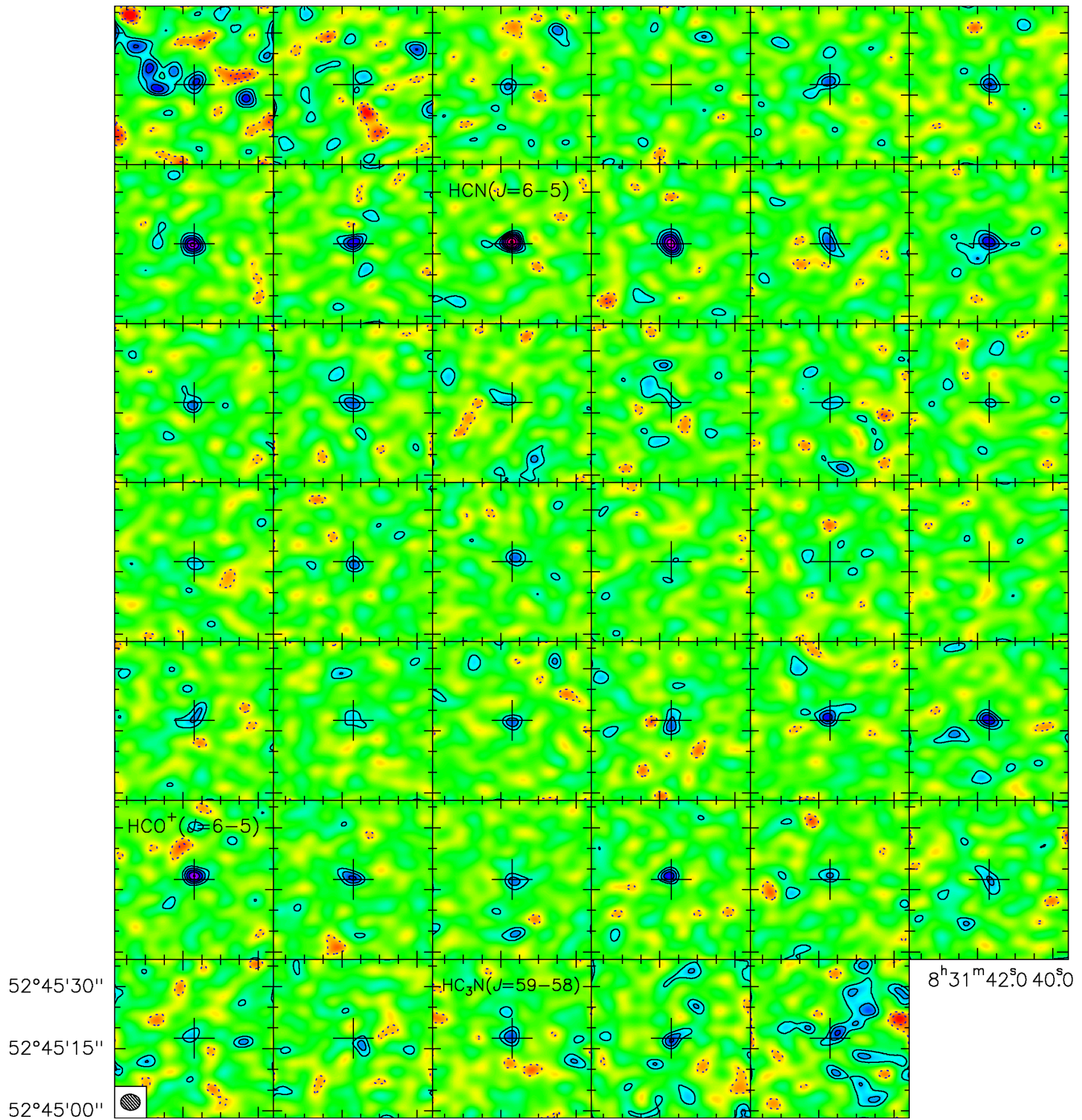}
%\vspace{-10mm}

\caption{Channel maps of the data shown as a spectrum in Fig.~\ref{f1} 
({\em left}) at the same velocity resolution. Frequencies increase
with channel number from left to right, i.e., red to blue. The peak
channels of the redshifted HCN, HCO$^+$, and HC$_3$N lines are
indicated. Contours are shown at (-3, -2, 2, 3, 4, 5, 6, 7,
8)\,$\sigma$ (1\,$\sigma$ = 0.95\,mJy\,beam$^{-1}$). The beam size
(4.5$''$$\times$3.6$''$) is shown in the bottom left corner. The cross
indicates a position of 08$^{\rm h}$31$^{\rm m}$41$^{\rm s}$.70,
+52$^\circ$45$'$17$''$.5 between the CO lens images (Riechers et al.\
\citeyear{rie09a}).
\label{f2}}
%\vspace{-5mm}
%
\end{figure*}

\subsection{Emission Lines of High-Density Gas Tracers}

\begin{figure*}
\epsscale{1.15}
\plotone{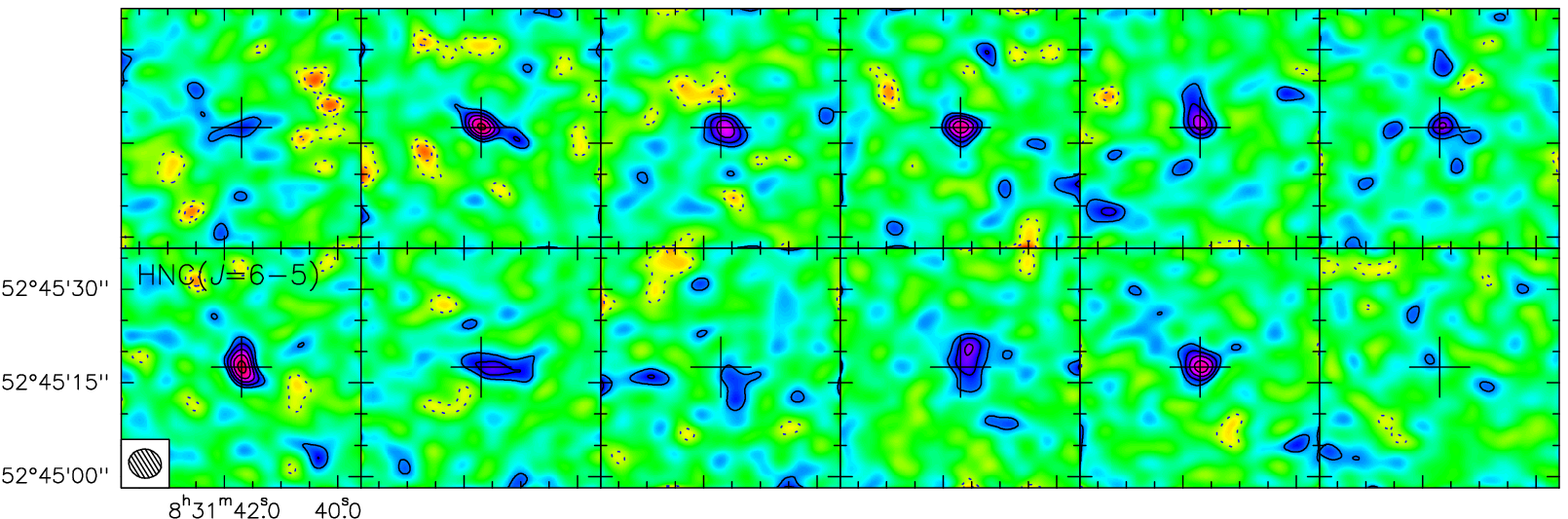}
%\vspace{-10mm}

\caption{Channel maps of the central 12 channels of the data shown as a 
spectrum in Fig.~\ref{f1} ({\em middle}) at the same velocity
resolution. Frequencies increase with channel number from left to
right, i.e., red to blue. The peak channel of the redshifted HNC line
is indicated. Contours are shown at (-3, -2, 2, 3, 4, 5, 6)\,$\sigma$
(1\,$\sigma$ = 0.88\,mJy\,beam$^{-1}$). The beam size
(5.3$''$$\times$4.6$''$) is shown in the bottom left corner. The cross
indicates the same position as in Fig.~\ref{f2}.
\label{f2a}}
%\vspace{-5mm}
%
\end{figure*}

We have detected emission from the $J$=6$\to$5 HCN, HCO$^+$ and HNC
emission lines toward the $z$=3.91 quasar \apm. The {\em left} panel
of Figure \ref{f1} shows the HCN and HCO$^+$ spectrum at 86\,\kms\
(31.25\,MHz) resolution. Both lines are detected simultaneously on top
of strong 2.76\,mm continuum emission.  The {\em middle} panel of
Figure \ref{f1} shows the HNC spectrum at similar (85\,\kms;
31.25\,MHz) resolution. The line is also detected on top of strong
continuum emission at similar wavelength (2.71\,mm). Figure \ref{f2}
shows the HCN and HCO$^+$ velocity channel maps at the same velocity
resolution and coverage as the spectrum in Fig.~\ref{f1}. At an rms of
0.95\,mJy\,beam$^{-1}$, \apm\ is detected at 8$\sigma$ and 6$\sigma$
significance in the peak channels of the HCN and HCO$^+$ lines.  Both
emission lines are detected over 10 channels on top of the continuum,
showing a decline in strength toward the linewings, as expected. Also,
marginal excess flux is detected at the peak position of the
redshifted \hcccn\ line, which however is not formally detected.
Figure \ref{f2a} shows the HNC velocity channel maps at the same
velocity resolution as the spectrum in Fig.~\ref{f1} (central 12
channels).  At an rms of 0.88\,mJy\,beam$^{-1}$, the source is
detected at 6$\sigma$ significance in the peak channel of the HNC
line, also showing a decline in flux towards the line wings.

From simultaneous Gaussian fitting to the profiles of the \fhcn\ and
\fhco\ lines and the underlying continuum emission, we derive HCN and
HCO$^+$ line peak flux densities of 4.04$\pm$0.61 and
2.48$\pm$0.56\,mJy and FWHM velocities of 385$\pm$36 and
385$\pm$60\,\kms, respectively (in good agreement with the
400$\pm$40\,\kms\ FWHM of \altehcn; Wei\ss\ et al.\ \citeyear{wei07}).
This corresponds to integrated HCN and HCO$^+$ line fluxes of
1.65$\pm$0.19 and 1.01$\pm$0.19\,Jy\,\kms. A simultaneous fit to the
\fhnc\ line and underlying continuum emission yields a HNC line peak
flux density of 1.57$\pm$0.51\,mJy and a line FWHM of
520$\pm$160\,\kms\ (also consistent with the FWHM of \altehcn\ within
the errors). This corresponds to an integrated HNC line flux of
0.86$\pm$0.24\,Jy\,\kms. From the line peak velocities, we derive
formal line redshifts of $z$=3.9126$\pm$0.0011 for HCN/HCO$^+$
(simultaneous fit), and $z$=3.9144$\pm$0.0011 for HNC, which are
consistent with those derived from other molecular emission lines
within the errors (e.g., Riechers et al.\ \citeyear{rie06b}; Wei\ss\
et al.\ \citeyear{wei07}).

We also searched for HC$_3$N $J$=57$\to$56, 59$\to$58, and 60$\to$59,
as well as \sio\ emission within the covered spectral range, including
the LSBs (combined spectrum shown in the {\em right} panel of Fig.\
\ref{f1}). Marginal excess flux is seen close to the peak position
of the redshifted \chcccn\ line in the overlap region of both LSB
setups, which we however consider not detected in the following. We
place 3$\sigma$ limits of 0.34, 0.52, 0.49, and 0.44\,Jy\,\kms\ on the
integrated fluxes from these lines, extracted over a fixed velocity
range of 400\,\kms\ (see Table \ref{t1}). We also attempted to stack
all three HC$_3$N lines, which results in no clear signal above the
formal 3$\sigma$ limit of 0.26\,Jy\,\kms. We however note that due to
the uncertainties in extraction, this `stacked' limit has to be
treated with caution.

\subsection{Millimeter Continuum Emission}

%%%%%%%%%%%%%%%%%%%%%%%%%%%%%%%%%%%%%%%%%%%%%%%%
%%%% Tab.2: Continuum Parameters
%%%%%%%%%%%%%%%%%%%%%%%%%%%%%%%%%%%%%%%%%%%%%%%%
\begin{deluxetable}{ l l l c}
\tabletypesize{\scriptsize}
\tablecaption{Measured continuum fluxes in \apm. \label{t2}}
\tablehead{
$\nu_{\rm obs}$ & $\lambda_{\rm obs}$ &  $\lambda_{\rm rest}$ & $S_{\nu}$ \\
${}$[GHz] & [mm] & [$\mu$m] & [mJy] }
\startdata
110.7 & 2.71 & 551 & 2.17$\pm$0.19 \\
108.6 & 2.76 & 562 & 2.08$\pm$0.22 \\
105.4\tablenotemark{a} & 2.84 & 579 & 2.04$\pm$0.08 \\
\hline
107.5\tablenotemark{b} & 2.79 & 568 & 2.08$\pm$0.07 
\vspace{-1mm}
\enddata 
\tablenotetext{a}{Combined LSB data from both setups.}
\tablenotetext{b}{Averaged over all continuum data.}
\end{deluxetable}

%%%%%%%%%%%%%%%%%%%%%%%%%%%%%%%%%%%%%%%%%%%%%%%%%%%%%%%%%%%

As mentioned above, we have detected $\sim$2.8\,mm (rest-frame
$\sim$560\,$\mu$m) continuum emission toward the host galaxy of the
$z$=3.91 quasar \apm. Emission was detected at high signal--to--noise
in each sideband of the two frequency setups. Thus, continuum fluxes
were extracted by fitting a two-dimensional, elliptical Gaussian to
the source in the $u-v$ plane for each sideband and frequency setting,
excluding ranges where line emission was detected. The individual
values are listed in Table~\ref{t2}, and are fully consistent with the
continuum fluxes obtained from the simultaneous line/continuum fits to
the spectra as outlined above. A combination of all measurements
yields an average continuum flux of 2.08$\pm$0.07\,mJy at 2.79\,mm,
consistent with the spectral energy distribution (SED) of the source
(Riechers et al.\ \citeyear{rie09a}).

\subsection{Line Luminosities and Ratios}

From the line intensities, we derive line luminosities and limits of
$L'_{\rm HCN(6-5)} = (4.9 \pm 0.6)$, $L'_{\rm HCO^+(6-5)} = (3.0 \pm
0.6)$, $L'_{\rm HNC(6-5)} = (2.4 \pm 0.7)$, $L'_{\rm HC_3N(57-56)} <
1.0$, $L'_{\rm HC_3N(59-58)} < 1.5$, $L'_{\rm HC_3N(60-59)} < 1.4$,
and $L'_{\rm SiO(12-11)} < 1.3
\times 10^{10}$\,$\mu_{\rm L}^{-1}$\,K\,\kms\,pc$^2$ (Tab.~\ref{t1}; not corrected for the lensing 
magnification factor of $\mu_{\rm L}$=4.2; Riechers et al.\
\citeyear{rie09a}). 

This corresponds to $J$=6$\to$5 $L'$ ratios of HCN/CO = 0.46$\pm$0.07,
HCO$^+$/CO = 0.28$\pm$0.06, and HNC/CO = 0.23$\pm$0.07 (relative to
\acoalt; Riechers et al.\ \citeyear{rie09a}).  Such high line ratios would 
be unusual even in the respective ground-state ($J$=1$\to$0)
transitions, and indicate exceptional dense gas excitation conditions
in this source.

We also find a HCN/HCO$^+$ $J$=6$\to$5 $L'$ ratio of 1.7$\pm$0.4,
which is higher than the HCN/HCO$^+$ $J$=5$\to$4 $L'$ ratio of
1.0$\pm$0.2 (Wei\ss\ et al.\ \citeyear{wei07}; Garcia-Burillo et al.\
\citeyear{gar06}). Such a steeply rising HCN/HCO$^+$ line ratio with $J$ is
not expected, as the densities and temperatures to collisionally
excite HCN and HCO$^+$ are similar. We note that the $J$=5$\to$4 lines
were not observed simultaneously, so the error bars of their ratio may
be underestimated due to the uncertainties in their relative flux
scales.  In addition, the difference in line ratio with $J$ may
indicate that \bhcccn\ line emission contributes significantly to the
\ehco\ flux reported by Garcia-Burillo et al., as these lines are
blended. The HC$_3$N limits derived above would be consistent with up
to a 30--40\% contamination (extrapolated from these higher-$J$
transitions) of the \ehco\ flux. Note that such high HC$_3$N/HCO$^+$
line ratios are actually observed in nearby infrared-luminous galaxies
(albeit in lower-$J$ lines; e.g., Aalto et al.\ \citeyear{aal07a}).

Furthermore, we find \fhnc/\fhcn\ = 0.50$\pm$0.15 and \fhnc/\fhco\ =
0.82$\pm$0.27. The \ehnc\ line was observed toward
\apm, but it is strongly blended with \dcn\ (Guelin et al.\ 
\citeyear{gue07}). The single Gaussian fit by Guelin et al.\ to the 
blend of these lines can be translated to upper limits of 
\ehnc/\ehcn\ $\leq$ 1.12 and
\ehnc/\ehco\ $\leq$ 1.15. Guelin et al.\ estimate that the CN line contributes 
$\sim$1/3 to the total flux they measure. Assuming their decomposition
is correct, we find \ehnc/\ehcn\ = 0.64$\pm$0.32 and \ehnc/\ehco\ =
0.66$\pm$0.33. Given the similar excitation densities and temperatures
of HNC compared to HCN and HCO$^+$, we assume that the $J$=5$\to$4 and
6$\to$5 line ratios should be comparable. This assumption is
consistent with the observed HNC/HCN ratios if the CN contribution to
the \ehnc\ $L'$ is as suggested by Guelin et al. It also is consistent
with the observed HNC/HCO$^+$ ratios, in particular if the HC$_3$N
contribution to \ehco\ is comparable to the CN contribution to \ehnc.
However, given the remaining uncertainties, observations of another CN
transition would be desirable to test this scenario (see also
discussion below).

Interestingly, we find a \fhcn/\ehcn\ $L'$ ratio of 1.36$\pm$0.31. A
ratio of greater than 1 is not expected for optically thick,
collisionally excited emission that is thermalized or
subthermal. However, such high $L'$ ratios have also been found for
HNC in nearby (ultra-) luminous infrared galaxies ((U)LIRGs), such as
Arp\,220 (albeit in lower-$J$ lines; Aalto et al.\
\citeyear{aal07b}). Moreover, extreme gas densities of $n$(H$_2$) $>
10^6$\,cm$^{-3}$ would be required to highly excite the $J$=6$\to$5
transitions of HCN, HNC and HCO$^+$ collisionally ($n_{\rm
crit}$(H$_2$) $> 10^8$\,cm$^{-3}$).

\section{HCN Line Excitation Modeling}

\begin{figure}
\epsscale{1.15}
\plotone{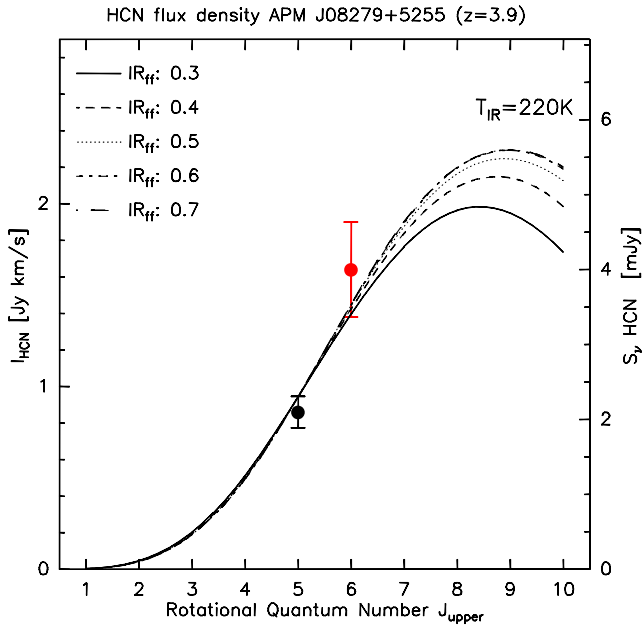}
%\vspace{-10mm}

\caption{HCN excitation ladder (spectral line energy distribution; 
points) and LVG models (lines) for \apm, accounting for both
collisional and radiative excitation.  The \ehcn\ data point is from
Wei\ss\ et al.\ (\citeyear{wei07}). The models give $n_{\rm
gas}$=10$^{4.2}$\,cm$^{-3}$, $T_{\rm kin}$=$T_{\rm IR}$=220\,K, and
infrared field filling factors of IR$_{\rm ff}$=0.3--0.7 (shown in
steps of 0.1). \label{f3}}
%\vspace{-5mm}
%
\end{figure}

To understand the unusual dense gas excitation in \apm\ in more
detail, we have carried out Large Velocity Gradient (LVG) models of
the HCN line excitation (see Wei\ss\ et al.\ \citeyear{wei07} for
initial study). Our models take both collisional and radiative
excitation via infrared (IR) pumping into account, using the HCN
collision rates from Sch\"oier et al.\ (\citeyear{sch05}), and
including the first ro-vibrational bending mode at 14.0\,$\mu$m for
pumping ($\nu_2=1$; e.g., Thorwirth et al.\ \citeyear{tho03}). We
adopted a HCN abundance per velocity gradient of [HCN]/(${\rm d}v/{\rm
d}r) = 1 \times 10^{-9}\,{\rm pc}\,$(\kms)$^{-1}$ (Helfer \& Blitz
\citeyear{hb97}; Wang et al.\ \citeyear{wan04}). All parameters are 
required to also reproduce the CO excitation of the source (Wei\ss\ et
al.\ \citeyear{wei07}), assuming [HCN/CO]=10$^{-4}$, and are required
to be consistent with the dust spectral energy distribution (Wei\ss\
et al.\ \citeyear{wei07}; Riechers et al.\ \citeyear{rie09a}). The
best solutions were obtained for a spherical, single-component model
with kinetic temperatures of $T_{\rm kin}$=220\,K, gas densities of
$n_{\rm gas}$=10$^{4.2}$\,cm$^{-3}$, an infrared radiation field
temperature of $T_{\rm IR}$(=$T_{\rm dust}$)=$T_{\rm kin}$=220\,K, and
infrared filling factors\footnote{IR$_{\rm ff}$ describes the solid
angle fraction of the gas that is exposed to the IR field, which in
turn is described by a grey-body spectrum.} of IR$_{\rm ff}$=0.3--0.7
(Fig.~\ref{f3}; where solutions with higher IR$_{\rm ff}$ prefer
slightly lower $T_{\rm kin}$ to stay consistent with the observed size
of the source in CO emission; Riechers et al.\ \citeyear{rie09a}).

Due to the high intensity of the IR-pumping field (as indicated by
IR$_{\rm ff}$) in these models, \ehcn\ and \fhcn\ are only moderately
optically thick (typical optical depths of $\tau_{5-4}$=0.25 and
$\tau_{6-5}$=0.4), and the optical depth still rises with $J$ between
these lines. This implies that the excitation of HCN is not
`super-thermal' (as in a `true' population inversion between energy
levels), but that the high HCN $L'$ ratio is merely an optical depth
effect. The HCN excitation is dominated by IR pumping for all models
that fit the data. Interestingly, the models imply that the HCN
emission cannot dominantly arise from a relatively cold, dense gas and
dust component (see Wei\ss\ et al.\ \citeyear{wei07} for
multi-component models). This means that standard conversion factors
$\alpha_{\rm HCN}=M_{\rm dense}/L'_{\rm HCN}$ from HCN line luminosity
to dense gas mass (e.g., Gao \& Solomon \citeyear{gs04}) will
substantially overpredict the dense molecular gas mass in this
system. Also, the models do not require unusually high relative HCN
abundances, contrary to previous suggestions based on a narrower
exploration of the parameter space in excitation conditions
(Garcia-Burillo et al.\ \citeyear{gar06}).

\section{Discussion}

\subsection{HCO$^+$ and HNC Excitation}

Due to the remaining uncertainties in the \ehco\ and \ehnc\ line
intensities (see discussion above), it is currently not possible to
properly constrain similar models for the HCO$^+$ and HNC line
excitation. However, due to similar ro-vibrational bending modes
($\nu_2=1$ at 12.1 and 21.6\,$\mu$m), it is likely that similar
conclusions hold for the HCO$^+$ and HNC excitation. The bending modes
of HCN, HCO$^+$ and HNC all lie close to the peak of the dust SED of
\apm\ (Wei\ss\ et al.\ \citeyear{wei07}; Riechers et al.\
\citeyear{rie09a}), which would make such a scenario plausible. Also,
the organic compound HC$_3$N has several bending modes in this
wavelength regime (e.g., Wyrowski et al.\ \citeyear{wyr99}), which
would make a substantial \bhcccn\ contribution to \ehco\
plausible. Luminous, high-$J$ HC$_3$N emission consistent with
radiative excitation scenarios is observed in nearby galaxies like
NGC\,4418 (Aalto et al.\ \citeyear{aal07a}).  On the other hand, the
lowest ro-vibrational transition of the diatomic molecule CN lies at
4.9\,$\mu$m (see also Guelin et al.\ \citeyear{gue07}), which lies
substantially beyond the peak of the SED. No strong 4.9\,$\mu$m
(observed-frame 24\,$\mu$m) CN absorption is detected towards the
bright mid-infrared continuum of \apm\ in deep {\em Spitzer Space
Telescope} spectroscopy (D.~Riechers et al., in prep.). Thus, IR
pumping is likely substantially less effective for CN (relative to
HCN, HCO$^+$ and HNC), which may contrast\footnote{The high CO
excitation in \apm\ can be explained by collisional excitation alone
(Wei\ss\ et al.\ \citeyear{wei07}); however, due to its higher
critical density, collisional excitation of CN is substantially less
effective than for CO.} the relatively high \dcn\ luminosity suggested
by Guelin et al.\ (\citeyear{gue07}), depending on the actual CN
abundance. However, luminous \ccn\ emission was detected toward
another distant lensed galaxy, the Cloverleaf quasar ($z$=2.56;
Riechers et al.\ \citeyear{rie07}).

If CN and/or HC$_3$N transitions close to those blended with
\ehnc\ and \ehco\ are found to be much fainter than expected based 
on the $L'$ estimates by Guelin et al.\ (CN) and above (HC$_3$N), a
more complex scenario would be required to explain the different HCN,
HCO$^+$, and HNC line ratios in the $J$=5$\to$4 and 6$\to$5
transitions.  In this case, it will become necessary to investigate
the different efficiency with which the IR radiation field excites
their bending modes in more detail, which may lead to high but
different excitation for the rotational ladders of these molecules.

\subsection{HNC/HCN Ratio}

Recent studies of HCN and HNC emission in nearby infrared-luminous
galaxies have revealed sources with high HNC/HCN ratios and high HNC
excitation. Two scenarios were brought forward to explain these high
ratios:\ IR pumping and/or increased abundances of HNC in the presence
of X-ray dominated regions (XDRs, often found in regions impacted by
emission from active galactic nuclei; Aalto et al.\
\citeyear{aal07b}). On the one hand, we find that \fhcn/\fhco\ $>$ 1,
which is the opposite to what is expected in the XDR scenario (Aalto
et al.\ \citeyear{aal07b}). However, we note that our LVG models show
that the \fhcn\ line (and thus, likely also the
\fhncalt\ line) is only moderately optically thick, abundance effects 
thus cannot be ruled out. On the other hand, the \fhnc/\fhcn\ ratio of
0.50$\pm$0.15 in \apm\ also fits well into the IR pumping scenario:\
in nearby infrared-luminous galaxies, the gas is warm enough to
efficiently pump HNC (through the 21.6\,$\mu$m bending mode with an
energy level of $h\nu/k$=669\,K and Einstein A coefficient of $A_{\rm
IR}$=5.2\,s$^{-1}$), but not HCN (through the 14.0\,$\mu$m bending
mode $h\nu/k$=1027\,K and $A_{\rm IR}$=1.7\,s$^{-1}$; Aalto et al.\
\citeyear{aal07b}), leading to a high HNC/HCN $L'$ ratio in high-$J$
transitions. In \apm, the dust and gas is warm enough to also
efficiently pump HCN at high rates, so the line ratio depends mostly
on the fraction of the HNC or HCN-emitting gas that is exposed to the
IR radiation field and optical depth effects (given that the fraction
of $L'$ in the $J$=6$\to$5 transitions due to collisional excitation
is likely small). This allows for a broad range of HNC/HCN ratios, and
thus is consistent with the comparatively low $J$=6$\to$5 ratio,
especially in combination with the high excitation of both molecular
line spectral energy distributions (SLEDs). Similar arguments can be
made for the relative strength of HCO$^+$, which has its fundamental
12.1\,$\mu$m bending mode at slightly higher energy than HCN.

\subsection{Detectability of Mir-IR Pumping Lines}

Given the SED and brightness of the high-$J$ rotational lines of HCN,
HNC, and HCO$^+$, pumping by mid-IR ro-vibrational transitions appears
plausible. As the abundance of these molecules is high enough to yield
bright rotational lines in emission, the question arises whether or
not the pumping transitions themselves may be detectable (which would
place better constraints on the abundance of these molecules). The
14.0\,$\mu$m HCN feature was detected in absorption in nearby (U)LIRGs
(Lahuis et al.\ \citeyear{lah07}). Indeed, sources like Arp\,220 or
NGC\,4418 show deep absorption features (10--30\% of the continuum
flux), indicating high HCN abundances, and evidence for pumping of the
rotational HCN lines.  The $\nu_2=1$ lines of HCO$^+$, HCN and HNC are
at rest-frame 12.1, 14.0, and 21.6\,$\mu$m, i.e., observed-frame 59.4,
68.8, and 106.1\,$\mu$m, which is within the wavelength range of the
Photodetector Array Camera and Spectrometer (PACS) on board the {\em
Herschel Space Observatory} (e.g., Poglitsch et al.\
\citeyear{pog10}). \apm\ has continuum fluxes of 511$\pm$51 and 
951$\pm$228\,mJy at 60 and 100\,$\mu$m (Irwin et al.\
\citeyear{irw98}). Thus, absorption features that have a depth of 10\%
of the continuum flux would be detectable within a few hours with {\em
Herschel}.

\section{Conclusions}

Based on CARMA observations of HCN, HNC, and \fhco\ toward the
$z$=3.91 quasar \apm, evidence is consolidating that the dense,
star-forming molecular gas in this source is substantially enhanced in
brightness by IR pumping (on top of gravitational magnification of the
source).\footnote{IR pumping may also explain the unusual excitation
of low-ionization rest-frame ultraviolet absorption lines in proximate
absorbers along selected sight lines toward the active galactic
nucleus (which may arise from diffuse gas in the galaxy; Srianand \&
Petitjean \citeyear{sp00}).} The unusual, hundreds of parsec scale,
warm gas and dust in this source appears to harbor a strong IR
radiation field that efficiently pumps the high-$J$ transitions of
HCN, HNC, and HCO$^+$ at rates well beyond those achieved by
collisional excitation alone. At least in the case of HCN, this leads
to moderately optically thick emission, which appears as
`super-thermal' between $J$=5$\to$4 and 6$\to$5 (due to a combination
of high excitation and rising optical depth). These findings are
consistent with the picture that the bulk of the gas and dust in this
source is situated in a compact, nuclear starburst, where both the
highly active galactic nucleus and star formation contribute to the
heating.

Even though \apm\ is known to be an extreme system, radiative
excitation of dense molecular gas tracers may also play a significant
role in other high redshift galaxies, like it does in nearby
infrared-luminous galaxies (e.g., Aalto et al.\
\citeyear{aal07b}). This issue becomes increasingly important for the
interpretation of observations of higher-$J$ ($J>2$) transitions of
dense gas tracers like HCN, HCO$^+$ and HNC, which are redshifted into
the millimeter observing windows at high $z$, and thus will be
primarily targeted toward galaxies in the early universe by future
facilities such as the Atacama Large Millimeter/submillimeter Array
(ALMA).

\acknowledgments 
We thank Christian Henkel for the original version of the LVG code.
DR acknowledges support from from NASA through Hubble Fellowship grant
HST-HF-51235.01 awarded by the Space Telescope Science Institute,
which is operated by the Association of Universities for Research in
Astronomy, Inc., for NASA, under contract NAS 5-26555.  Support for
CARMA construction was derived from the G.\ and B.\ Moore Foundation,
the K.~T.\ and E.~L.\ Norris Foundation, the Associates of the
California Institute of Technology, the states of California,
Illinois, and Maryland, and the NSF. Ongoing CARMA development and
operations are supported by the NSF under a cooperative agreement, and
by the CARMA partner universities.

\appendix

\section{Flux Calibration}

\begin{figure}
\epsscale{0.61}
\plotone{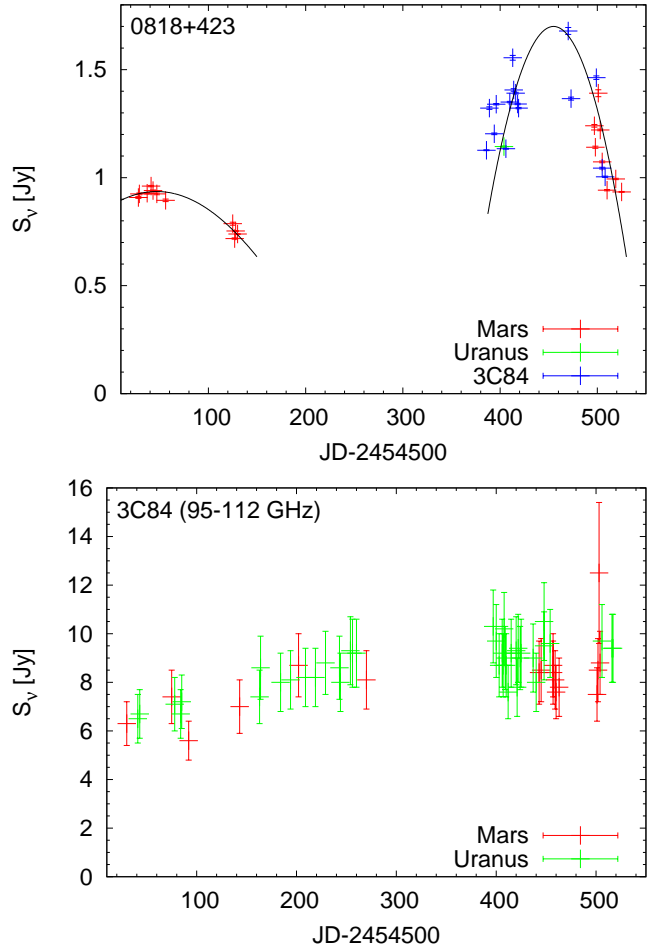}
%\vspace{-10mm}

\caption{Millimeter variability of the phase calibrator J0818+423. 
{\em Top:} Measured $\sim$105\,GHz flux for the tracks in this project
where a primary calibrator was observed, bootstrapped relative to the
flux calibrators indicated in the bottom right corner. Error bars
indicate the variance of individual measurements within a track. The
black line indicates a fit to the data. {\em Bottom:} Measured
95--112\,GHz fluxes of 3C84 over the same time range (CARMA archive;
S.~Schnee, priv.\ comm.), bootstrapped relative to the flux
calibrators indicated in the bottom right corner. Error bars include
the absolute calibration uncertainty. This comparison shows that the
variability among tracks bootstrapped relative to 3C84 is not due to
intrinsic short-term variability of 3C84 itself.
\label{f6}}
%\vspace{-5mm}
%
\end{figure}

The observations presented here were taken over the course of four
(HCN \& HCO$^+$)/six (HNC) months. It thus is necessary to correct the
flux scale of individual tracks for the intrinsic variability of the
phase calibrator. Given the relatively large number of tracks (14/23),
this leads to a substantially improved flux calibration relative to
the sparsely sampled, internal look-up table of MIRIAD.

Figure \ref{f6} shows the variability of J0818+423, the phase
calibrator, bootstrapped relative to Mars, Uranus and 3C84 ({\em top
panel}), measured over the first band in the LSB at
$\sim$104.6/105.3\,GHz. For illustration, a fit to the data is shown.
Fluxes for tracks that were observed without a primary calibrator were
estimated based on a weighted average of those of tracks that are
adjacent in time.  As 3C84 is a millimeter-variable source
itself,\footnote{See, e.g., SMA data archive; {\tt
http://sma1.sma.hawaii.edu/callist/} .} the {\em bottom} panel shows
its variability over the same time range in which this project was
observed. Within the calibration errors, the phase calibrator shows a
clear excess variability over that of 3C84 in the 2009 measurements
(data points at (JD--2454500)$>$332). Also, the scatter of values in
the 2009 measurements bootstrapped relative to the Mars planetary
model is substantially larger than that in 2008 (typically observed
under comparable weather conditions). Together with the fact that the
fluxes bootstrapped relative to 3C84 agree well with those
bootstrapped relative to planet models, this suggests that J0818+423
shows significant intrinsic flux variations over the observed time
range. Thus, we obtain the best calibration by using and interpolating
the individually bootstrapped fluxes for each track, rather than using
an averaged value for J0818+423. We conservatively estimate that this
flux calibration leads to an overall accuracy of $\sim$15\%.


\begin{thebibliography}{}

\bibitem[Aalto et al.(2007a)]{aal07a} Aalto, S., Monje, R., \& Mart\'in, 
  S.\ 2007a, A\&A, 475, 479
\bibitem[Aalto et al.(2007b)]{aal07b} Aalto, S., Spaans, M., Wiedner, M.~C., 
  \& H\"uttemeister, S.\ 2007b, A\&A, 464, 193
\bibitem[Baan et al.(2008)]{baa08} Baan, W.~A., Henkel, C., Loenen, A.~F., 
  Baudry, A., \& Wiklind, T.\ 2008, A\&A, 477, 747
\bibitem[Barvainis et al.(1997)]{bar97} Barvainis, R., Maloney, P.,
  Antonucci, R., \& Alloin, D.\ 1997, ApJ, 484, 695
\bibitem[Gao \& Solomon(2004)]{gs04} Gao, Y., \& Solomon, P.~M.\
  2004, ApJ, 606, 271
\bibitem[Gao et al.(2007)]{gao07} Gao, Y., Carilli, C.~L., Solomon,
  P.~M., \& Vanden Bout, P.~A.\ 2007, ApJ, 660, L93
\bibitem[Garc\'ia-Burillo et al.(2006)]{gar06} Garc\'ia-Burillo, S.,
  Graci\'a-Carpio, J., Gu\'elin, M., Neri, R., Cox, P., Planesas, P.,
  Solomon, P.~M., Tacconi, L.~J., \& Vanden Bout, P.~A.\ 2006, ApJ,
  645, L17
\bibitem[Graci\'a-Carpio et al.(2008)]{gra08} Graci\'a-Carpio, J., et al.\ 2008, A\&A, 479, 703
\bibitem[Gu\'elin et al.(2007)]{gue07} Gu\'elin, M., et al.\ 2007, A\&A, 
  462, L45
\bibitem[Helfer \& Blitz(1997)]{hb97} Helfer, T., \& Blitz, L.\ 1997, 
  ApJ, 478, 233
\bibitem[Irwin et al.(1998)]{irw98} Irwin, M.~J., \etal\ 1998, ApJ, 505, 529
\bibitem[Lahuis et al.(2007)]{lah07} Lahuis, F., \etal\ 2007, ApJ, 659, 296
\bibitem[Poglitsch et al.(2010)]{pog10} Poglitsch, A., \etal\ 2010, A\&A, 518, L2
\bibitem[Riechers et al.(2006a)]{rie06a} Riechers, D.~A., Walter, F.,
  Carilli, C.~L., \etal\ 2006a, ApJ, 645, L13
\bibitem[Riechers et al.(2006b)]{rie06b} Riechers, D.~A., Walter, F.,
  Carilli, C.~L., \etal\ 2006b, ApJ, 650, 604
\bibitem[Riechers et al.(2007a)]{rie07a} Riechers, D.~A., Walter, F., 
  Carilli, C.~L., \& Bertoldi, F.\ 2007a, ApJ, 671, L13
\bibitem[Riechers et al.(2007b)]{rie07} Riechers, D.~A., Walter, F.,
  Cox, P., \etal\ 2007b, ApJ, 666, 778
\bibitem[Riechers et al.(2009)]{rie09a} Riechers, D.~A., Walter, F., 
  Carilli, C.~L., \& Lewis, G.~F.\ 2009, ApJ, 690, 463
\bibitem[Schoier et al.(2005)]{sch05} Sch\"oier, F.~L., van der Tak, 
  F.~F.~S., van Dishoeck, E.~F., \& Black, J.~H.\ 2005, A\&A, 432, 369
\bibitem[Solomon et al.(1992)]{sol92} Solomon, P.~M., Radford,
  S.~J.~E., \& Downes, D.\ 1992, Nature, 356, 318
\bibitem[Solomon et al.(2003)]{sol03} Solomon, P., Vanden Bout, P.,
  Carilli, C., \& Guelin, M.\ 2003, Nature 426, 636
\bibitem[Solomon \& Vanden Bout(2005)]{sv05} Solomon, P.~M., \& Vanden
  Bout, P.~A.\ 2005, ARA\&A, 43, 677
\bibitem[Spergel et al.(2003)]{spe03} Spergel, D.~N., Verde, L.,
  Peiris, H.~V., \etal\ 2003, ApJS, 148, 175
\bibitem[Spergel et al.(2007)]{spe07} Spergel, D.~N., Bean, R.,
  Dor\'e, O., \etal\ 2007, ApJS, 170, 377
\bibitem[Srianand et al.(2000)]{sp00} Srianand, R., \& Petitjean, P.\ 2000, 
 A\&A, 357, 414
\bibitem[Thorwirth et al.(2003)]{tho03} Thorwirth, S., Wyrowski, F., 
  Schilke, P., Menten, K.~M., Br\"unken, S., M\"uller, H.~S.~P., \&
  Winnewisser, G.\ 2003, ApJ, 586, 338
\bibitem[Wagg et al.(2005)]{wag05} Wagg, J., Wilner, D.~J., Neri, R.,
  Downes, D., \& Wiklind, T.\ 2005, ApJ, 634, L13
\bibitem[Wang et al.(2004)]{wan04} Wang, M., Henkel, C., Chin, Y.-N., 
  et al.\ 2004, A\&A, 422, 883
\bibitem[Wei\ss\ et al.(2007)]{wei07} Wei\ss, A., Downes, D., Neri,
  R., \etal\ 2007, A\&A, 467, 955
\bibitem[Wyrowski et al.(1999)]{wyr99} Wyrowski, F., Schilke, P., \& 
  Walmsley, C.~M.\ 1999, A\&A, 341, 882
\end{thebibliography}
\end{document}